\documentclass[fleqn,usenatbib]{mnras}

\usepackage[T1]{fontenc}

\DeclareRobustCommand{\VAN}[3]{#2}
\let\VANthebibliography\thebibliography
\def\thebibliography{\DeclareRobustCommand{\VAN}[3]{##3}\VANthebibliography}

\usepackage{graphicx}	
\usepackage{amsmath}	
\usepackage{amssymb}	

\usepackage{newtxtext,newtxmath}

\title[An optical and X-ray faint CV?]{UGPS~J194310+183851: an Unusual Optical and X-ray Faint Cataclysmic Variable?}

\author[C. Morris et al.]{
C. Morris$^{1}$\thanks{E-mail: c.morris6@herts.ac.uk},
T. J. Maccarone$^{2}$,
P. W. Lucas$^{1}$,
J. Strader$^{3}$,
C. T. Britt$^{4}$, 
\newauthor
 N. Miller$^{1}$,
S. J. Swihart$^{5}$,
W. J. Cooper$^{1,6}$,
J. E. Drew$^{1,7}$,
and Z. Guo$^{8,9,1}$
\\
$^{1}$Centre for Astrophysics Research, University of Hertfordshire, College Lane, Hatfield, Hertfordshire, AL10 9AB, UK\\\
$^{2}$Department of Physics \& Astronomy, Texas Tech University, Box 41051, Lubbock, TX 79409-1051, USA\\
$^{3}$Center for Data Intensive and Time Domain Astronomy, Department of Physics and Astronomy, Michigan State University, East Lansing, MI 48824, USA\\
$^{4}$Space Telescope Science Institute, 3700 San Martin Dr, Baltimore, MD 21218, USA\\
$^{5}$National Research Council Research Associate, National Academy of Sciences, Washington, DC 20001, USA,\\
Resident at Naval Research Laboratory, Washington, DC 20375, USA\\
$^{6}$Istituto Nazionale di Astrofisica, Osservatorio Astrofisico di Torino, Strada Osservatorio 20, I-10025 Pino Torinese, IT\\
$^{7}$Department of Physics \& Astronomy, University College London, Gower Street, London WC1E 6BT, UK\\
$^{8}$Instituto de F{\'i}sica y Astronom{\'i}a, Universidad de Valpara{\'i}so, ave. Gran Breta{\~n}a, 1111, Casilla 5030, Valpara{\'i}so, Chile\\
$^{9}$N\'ucleo Milenio de Formaci\'on Planetaria (NPF), Casilla 5030, Chile\\
}

\date{Accepted XXX. Received YYY; in original form ZZZ}

\pubyear{2015}

\begin{document}

\newpage
\newpage

\label{firstpage}
\pagerange{\pageref{firstpage}--\pageref{lastpage}}
\maketitle

\begin{abstract}

The growing number of multi-epoch optical and infrared sky surveys are uncovering unprecedented numbers of new variable stars, of an increasing number of types. The short interval between observations in adjacent near infrared filters in the UKIDSS Galactic Plane Survey (UGPS) allows for the discovery of variability on the timescale of minutes. 
We report on the nature of one such object, through the use of optical spectroscopy, time-series photometry and targeted X-ray observations. We propose that UGPS J194310.32+183851.8 is a magnetic cataclysmic variable star of novel character, probably featuring a longer than average spin period and an orbital period likely to be shorter than the period gap (i.e. $\text{P}_{\rm{orb}}<$2 hours). We reason that the star is likely a member of the short period Intermediate-Polar subclass that exist below this period boundary, but with the additional feature that system's SED is fainter and redder than other members of the group.

\end{abstract}
 



\section{Introduction}



The object UGPS J194310.32+183851.8 (hereafter referred to as Source 363, \citealt{Lucas2017}) was imaged twice as part of the UKIDSS Galactic-Plane Survey (UGPS), displaying 1.97 mag infrared variability in K over the 4 year interval and H$\alpha$ emission in the IPHAS survey photometry \citep{drew05}. One of the infrared colours was exceptionally blue: $J-H = -0.6\text{, but }H-K = +0.7$. Lucas et al. suggested that this might be due to variability on the 7 minute timescale of UKIDSS filter changes, supported by marginal detections in the 2MASS images that suggested unremarkable colours. Additionally, the Gaia satellite has determined a fairly nearby location (see Section 2.1). Consequently, from the lack of any previous detection of hard X-rays, gamma rays or radio waves by facilities such as Fermi-LAT, ROSAT and  the Very Large Array Sky Survey (VLASS), \citep{fermiLAT,ROSAT,Lacy2020}, we infer that the system does not contain a neutron star or black hole. Thus we must be looking at a white dwarf (WD) binary system most likely of the cataclysmic variable (CV) type.

Cataclysmic variables (CVs) are binary systems containing a WD and a close companion (of comparatively low mass), wherein the WD primary accretes mass via Roche-lobe overflow from the secondary star (see \citealt{Knigge2011} and  \citealt{Schreiber2021}  for comprehensive reviews). These systems are subdivided into magnetized and unmagnetized systems (based largely on the presence or absence of pulsations caused by the accretion flow proceeding along the magnetic field lines), and further subdivided based upon the method of WD accretion.  The magnetic CVs consist of polars and intermediate polars (IPs). Polars accrete matter following the magnetic field lines directly to the surface of the primary due to their high magnetic field strengths of 7-230MG \citep{Oliveira2017}. This high field strength also causes the orbit and the spin of the WD's magnetosphere to become synchronous. In contrast, in IPs the Alfven radius is smaller than the circularization radius of the accretion disc, but larger than the white dwarf's radius, allowing an outer disc to exist. This disc is truncated, and then leads to the flow following magnetic field lines at free-fall velocity in the inner region. These systems are also notable for the spin and orbital periods ($P_{\mathrm{spin}}$ \& $P_{\mathrm{orb}}$)  being different from one another. 
It should be noted that a small subgroup of IPs can transfer mass via both disc-fed and stream-fed methods; for example FO Aqr, TX Col (\citealt{Littlefield2020}, \citealt{Littlefield2021}) and others still are known to be solely stream-fed and are referred to as discless IPs, such as EX Hya and DW Cnc \citep{2013Andronov,Rodriguez-Gil2004}.

Their hard spectrum arises because the gas is in free-fall on to a white dwarf, so if the accretion column is optically thin, the emission will come out as bremsstrahlung radiation with energies $E \approx kT \sim 10-30$~keV.  The innermost region will usually be optically thick, and then can give emission in soft X-rays or EUV, depending on the luminosity and the size scale of the polar region spot at the base of the accretion column. It's reasonable that some of these are missed because of absorption effects, but that's still unsettled. X-ray spectra frequently show ionized iron in emission, caused by shock excitation from the free-falling disk material.

In the optical regime, the systems are highly variable and multi-periodic, with periodicities associated with the WD spin, the binary orbit and beat interactions between these two periods. Spectrally, IPs are characterised by broad emission of hydrogen recombination lines, as well as emission of helium (both neutral and ionized), carbon and nitrogen. Absorption features can be used to understand the composition of the companion (providing they are of high signal to noise), as well as the disc inclination. By comparison, polars have stronger but narrower emission features (often having Zeeman splitting due to the high magnetic field strength), including a prominent doubly ionized nitrogen line at 4650$\text{\AA}$ produced in the accretion column.  

In this paper we present the discovery of Source 363's nature as a new magnetic CV. We report on optical spectroscopic observations showing unusually broad emission lines (Section 3.1). We then detail numerous photometric monitoring observations which illustrate the system's behaviour on short timescales (Section 3.2), revealing periodicity at $\approx$ 2760s. Finally we report on new X-ray observations by the {\it Neil Gehrels Swift Observatory} (hereinafter {\it Swift}) which have provided the first detections of X-ray emission from the source, consistent with our classification of the source as an accreting WD system (Section 3.3). We then conclude with a discussion on the nature and novel properties of the source, given the various dichotomies our combined approach has uncovered (Section 4).

\section{Source Information and Observations}

\subsection{Source Characteristics} \label{Source_info}

Previous optical and infrared observations of the source have been carried out as part of UGPS, 2MASS, UVEX, IPHAS, Gaia and Pan-STARRS (\citealt{Lucas2008, 2mass, uvex, drew05, gaiadr2, gaia_edr3_2, pannstars1}). They reveal the source to be an optically faint but comparatively infrared-bright variable star (Table 1\footnote{[1]:\cite{2014CP_UKIDSS_I,Lucas2008},[2]:\cite{IGAPS_2020},[3]:\cite{gaia_edr3_2,gaiadr2,gaia2016},[4]:\cite{pannstars1}})
, with a prominent H$\alpha$~excess. Rapid variability is clearly seen in the sparsely sampled Pan-STARRS 1 optical light curve: it varies by a factor of 4 in 15 minutes or less in $r$, confirming the fast nature of the source's variability. This indicates that the system includes a compact object of some type. The latest distance from \cite{Bailer_Jones2021} using Gaia EDR3 put the star at $900.7 \substack{+288.3 \\ -245.6}$~pc, which leads to an absolute magnitude in $G$ of 10.61, although some care should be taken with this due to the variability of the source, as the prior used by \citet{Bailer_Jones2021} is more suited to non-variable stars. All these features together were indicative of Source 363 possibly being a compact object, although there was no associated X-ray emission recorded in previous all-sky surveys, so the flux must be less than $5\times10^{-13}\text{erg\ cm}^{-2}\ \text{s}^{-1}$, the limit for ROSAT \citep{ROSAT}. The absence of X-rays of the required intensity can safely rule out a neutron star system, but a CV system would still be plausible for a faint X-ray source. The combination of the strong, rapid variability and the lack of X-ray emission in all-sky surveys led to the acquisition of targeted X-ray observations, several high cadence light curves, and an optical spectrum.        

\begin{figure}
	\includegraphics[width=\columnwidth]{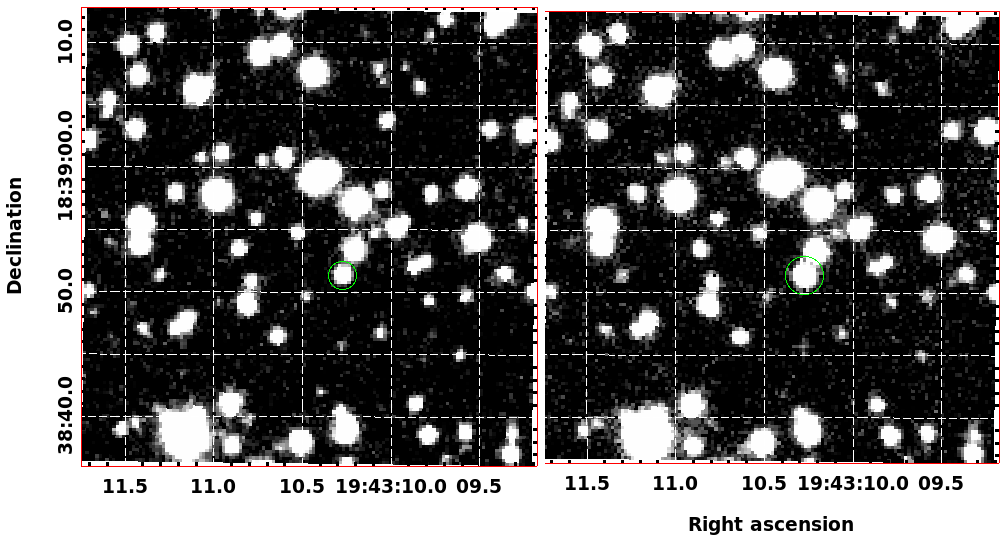}
    \caption{WHT PF-QHY r'-band images of Source 363 (within the green annulus), showing the short timescale variation, with the 1.1 magnitude change across 16 minutes. The left panel has an MJD of 59386.153 and the right panel of 59386.164.}
    \label{fig:wht_rvar}
\end{figure}

\begin{table*}
	\centering
	\caption{Table of selected archival measurements for Source 363. N.B., an asterisk in No. of Epochs implies that more epochs were available but they are not included because of poor data quality.}
	\label{tab:arc_data_tab}
    \begin{tabular}{lcccr}
        \hline
        Survey & Filter & Mean Magnitude \& Error & No. of Epochs & Maximum Variation (mag) \\
        \hline
        UKIDSS$^{\text{[1]}}$ & $Ks$ & 16.70$\pm$0.08 & 2 & 1.97 \\
        UKIDSS$^{\text{[1]}}$ & $H$ & 18.36$\pm$0.11 & 1 & N/A \\
        UKIDSS$^{\text{[1]}}$ & $J$ & 17.76$\pm$0.03 & 1 & N/A \\
        IGAPS$^{\text{[2]}}$ & $r$ & 20.54$\pm$0.06 & 4 & 1.94 \\
        IGAPS$^{\text{[2]}}$ & $H\alpha$ & 19.42$\pm$0.06 & 2 & N/A \\
        IGAPS$^{\text{[2]}}$ & $i$ & 19.27$\pm$0.07 & 2 & N/A \\
        IGAPS$^{\text{[2]}}$ & RGO $U$ & 19.68$\pm$0.03 & 2 & 0.37 \\
        Gaia$^{\text{[3]}}$ & $G$ & 20.02$\pm$0.10 & 3 & 0.72 \\
        Pan-STARRS$^{\text{[4]}}$ & $g$ & 20.35$\pm$0.04 & 8* & 1.46 \\
        Pan-STARRS$^{\text{[4]}}$ & $r$ & 20.83$\pm$0.07 & 8 & 1.82 \\
        Pan-STARRS$^{\text{[4]}}$ & $z$ & 19.64$\pm$0.07 & 4* & 0.55 \\
        \hline
    \end{tabular}
\end{table*}

\subsection{SOAR Observations}

\subsubsection{Spectroscopy}

We obtained a single low-resolution spectrum of the source on 2021 May 12 with the Red Camera of the Goodman Spectrograph \citep{Clemens2004} on the SOAR telescope.
The exposure time was 1800 s, and the observation used a 400 l mm$^{-1}$ grating and a 0.95\arcsec slit together with a GG395 long-pass filter, yielding a full-width at half-maximum resolution of 5.4$\text{\AA}$ (about 250 km s$^{-1}$) over a usable wavelength range $\sim 4000$--7820 \AA.

The spectrum was reduced and optimally extracted in the usual manner using IRAF \citep{Tody1986}. The resulting spectrum has an average signal to noise ratio of $\sim 14$ per resolution element ($\sim 8$ per pixel) in the continuum.

\subsubsection{Imaging}

Follow up imaging was obtained on 2021 June 9, using the Red Camera of SOAR/Goodman in imaging mode. We obtained 29 exposures, each of 180~s duration, using the SDSS $i'$ filter. These totalled 87 min on source over a time span of 98 min. The average airmass was 1.62 and typical seeing was $1.5\arcsec$.

The raw images were corrected for bias and then flat-fielded with sky flats using standard routines in IRAF \citep{Tody1986}. We performed differential photometry of Source 363 with respect to 17 nearby, non-variable stars and calibrated these magnitudes using their Pan-STARRS DR2 \citep{PS1DR2} $i'$ mags, the final results can be seen in Figure \ref{fig:curve_I}.

\subsection{WHT Observations}

We made use of recently offered service time on the William Herschel Telescope for testing the new PF-QHY camera\footnote{See the instrument page for details: 'www.ing.iac.es/astronomy/instruments/pf-qhy/'} to obtain two high cadence light curves. These were using Sloan u' and r' filters, with 62 \& 61 epochs of data covering 90 and 89 minutes respectively and individual exposures of 80 seconds in each band. PF-QHY is  a wide-field camera with a CMOS detector, its field of view being 10.7' by 7.1' and a pixel scale of $0.267"$ using 4x4 binning. Data reduction and calibration were carried out using the \texttt{photutils} \citep{larry_bradley_2021} and \texttt{astropy} \citep{astropy1,astropy2} libraries for \texttt{python}, with profile-fitting photometry being used in the crowded r' images.  Aperture photometry was used in the u' images as the field is comparatively less  populated with sources. Calibration was done via a standard star SP~1942+261, with corrections to individual epochs being averaged across non-variable stars in each frame.  

\subsection{Swift XRT/UVOT Observations} \label{swift_obs}

We obtained X-ray and UV data from Swift via a successful target of opportunity (ToO) program on the 2$^{\text{nd}}$ of July 2021, using both the XRT and UVOT instruments in parallel, with 4 ks of observation in the photon-counting mode. These data produced the standard X-ray images, light curve and spectrum, as well as additional images and a light curve in the UVOT U filter. The data were provided pre-reduced via the Swift ToO server (\citealt{SwiftDetect,SwiftMethods}) with the spectrum being fitted with a power law, and adjusted for galactic absorption along the line of sight. Calibration for the UVOT data was done in \texttt{python} using the same packages described previously. All other XRT reduction was carried out using \texttt{XSPEC}, using the specific setup for Swift-XRT.    

\section{Results and Analysis}
\subsection{Optical Spectrum}
\begin{figure*}
	\includegraphics[width=\linewidth]{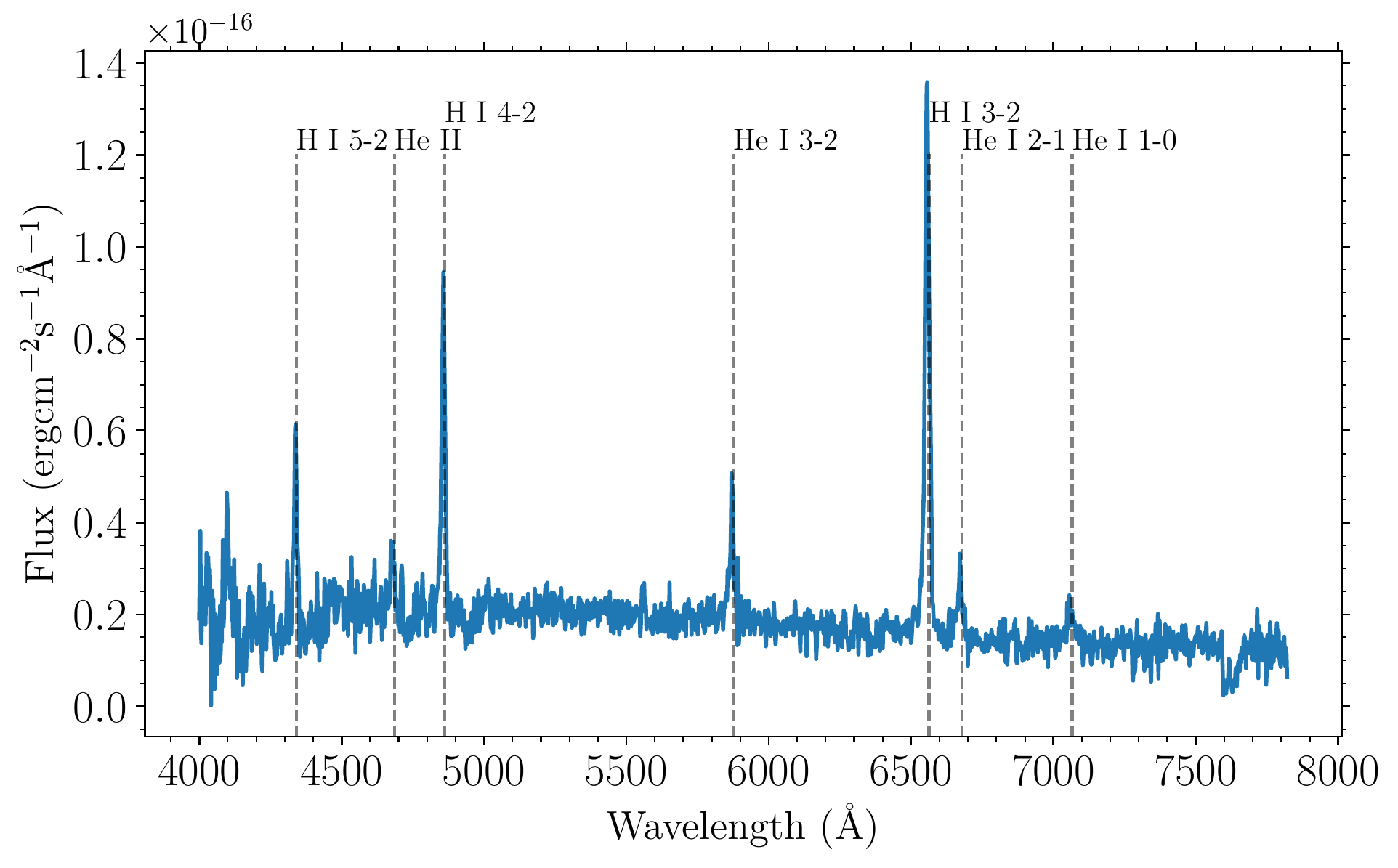}
    \caption{Optical spectrum of Source 363, using the Goodman Spectrograph at SOAR. Hydrogen recombination lines are prominent, and broader than those seen in more typical CVs. Other common CV lines such as \ion{He}{i}and \ion{He}{ii}are also present, although the 4606\AA $\,$ \ion{Na}{iii} line often associated with IPs is not seen above the noise.}
    \label{fig:soar_spec}
\end{figure*}

\begin{figure}
	\includegraphics[width=\linewidth]{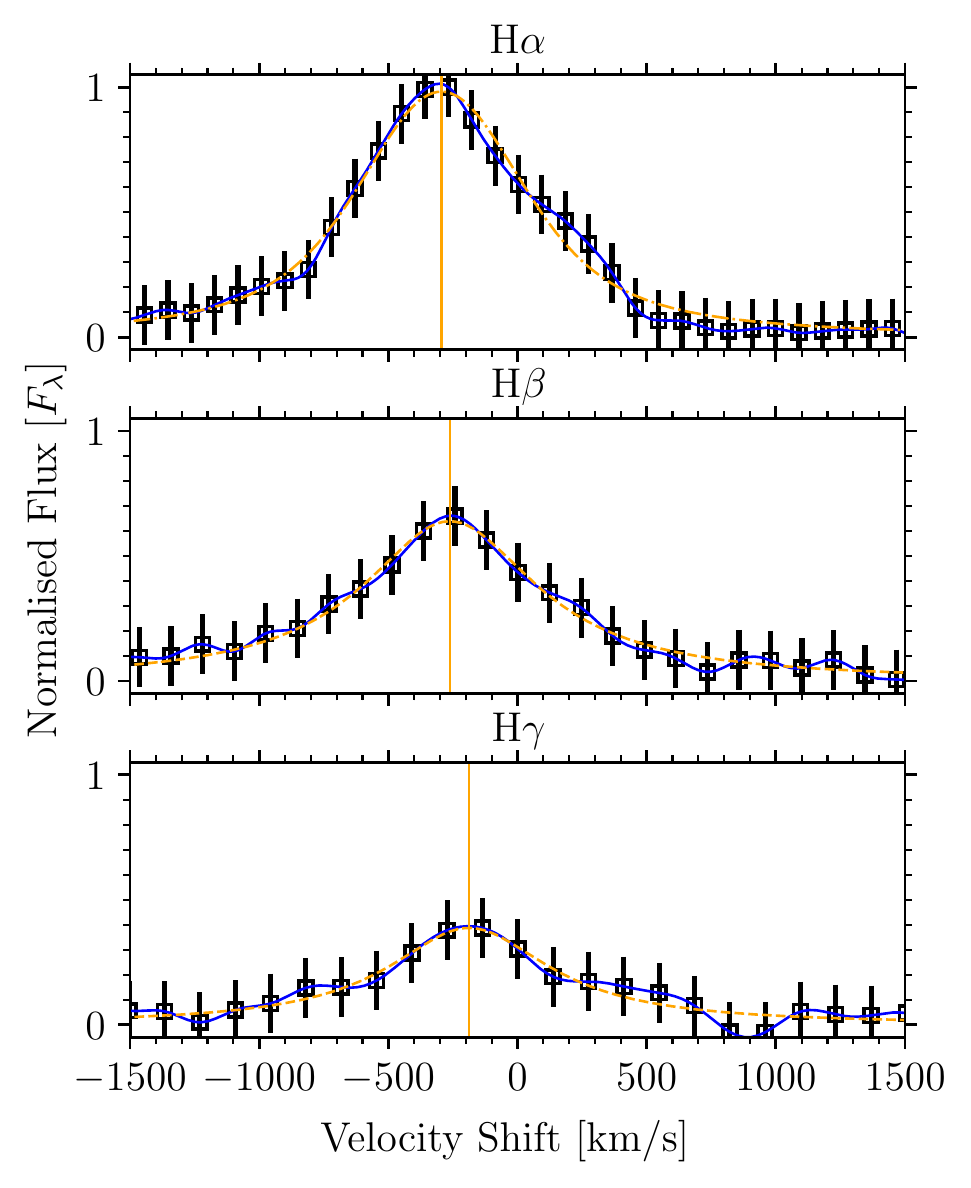}
    \caption{
    Fitted line profiles and radial velocities for the hydrogen recombination lines. Points and blue lines are the data, yellow dashed lines are the fits, using a Voigt profile, and converted into velocity space.
    The median radial velocity of the star was $-283.9\pm9.3$~km$s^{-1}$, relative to the Local Standard of Rest. The individual radial velocities fitted to each line are marked with solid yellow lines. 
    Fluxes are normalised with respect to the HI $3\rightarrow2$ line.
    }
    \label{fig:soar_spec_rvs}
\end{figure}

The SOAR spectrum (Figure \ref{fig:soar_spec}) displays a number of features consistent with CVs in general, most notably H$\alpha$ and H$\beta$ in emission, with additional emission features from both neutral and singly-ionized helium (see Table \ref{tab:ew_tab} for details of individual lines). All emission lines are broad and single-peaked, with a wavelength resolution of $100.5\ \text{kms}^{-1}$ at 5910$\text{\AA}$ (the mid-point of the spectrum). The wide range of velocities suggests the presence of a disc, or possibly a symmetric wind, although this cannot be further identified due to the blending of any \ion{C}{iii} / \ion{N}{iii} lines into the wings of the \ion{He}{ii} line at 4686\AA. Fitting of line profiles (using a Voigt profile to account for both Doppler and Stark broadening) provides an average radial velocity (RV) of $-280.2\pm12.9 \text{kms}^{-1}$, see Figure \ref{fig:soar_spec_rvs}. This value, while high, is not unexpected when only one spectrum is available, and the system's mean radial velocity is likely less than this. Some absorption features could be present in the spectrum, however the modest S/N ($\approx$14 per resolution element) precludes any reliable identification of these, and thus we have to determine the characteristics of the companion star through another method.   

From the equivalent width (EW) ratio of ${\ion{He}{ii}/\mathrm{H{\beta}}} = 0.18\pm0.10$ we cannot be certain whether the system's accretion is primarily magnetic (see \citealt{Silber1992}), due to the noise in the detection of the \ion{He}{ii} line: a ratio above 0.4 is indicative of magnetically controlled accretion. It should be noted that those authors also suggest that the EW of H$\beta$ should be $\geq 20$\AA ~in magnetic systems: here we have EW=49.3\AA. We reason that the observed hydrogen line ratios may be as a result of a more complete than usual disk (for an IP), bearing resemblance to those observed around dwarf-novae. In those cases the slower rotating outer disk regions limit the excitation of the more energetic recombination lines, leading to lower fluxes from H$\beta$ and H$\gamma$ than normal IPs.       

We cannot be sure if Source 363 was in a brighter than average state at the time of observation; pre-spectroscopic imaging places the star at an intermediate brightness level (compared to its maxima and minima). However we cannot interpret whether the star was pre- or post-outburst, which given the length of the subsequent observation could place the star at either state. This is especially true given the low S/N of the \ion{He}{ii} line, meaning that the true ratio might be higher than the 0.4 value that is assumed to indicate magnetized accretion. 

In comparison to objects from the literature, we note the absence of emission from \ion{N}{iii} and \ion{C}{iii}, which is commonly seen in fully magnetic polars \citep{warner_1995}. In addition the fringes of the $H_{\beta}$ line do not contain the characteristic absorption features that are associated with a high disc inclination angle. Given that a broad range of velocities is observed the disc inclination angle cannot be very low so we can reason that the inclination angle is intermediate.
The same line can also show the hallmarks of cyclotron radiation, as an overall increase in the localised continuum, which we note that Source 363 does not display, so cyclotron radiation is unlikely to be the cause of the observed near infrared excess.

\begin{table}
	\centering
	\caption{Table of equivalent line widths for notable features in the I-band spectrum. While faint absorption features from a companion star can possibly be seen, they are not prominent enough for us to provide analysis. It's worth noting the approximate 9/3/1 ratio between the Balmer lines, while the helium ground state transitions indicate a higher population at higher energy levels, with the 3-2 line being most prominent. The helium lines were affected by noise and some blending, and thus have not been fitted for RVs and widths.}
	\label{tab:ew_tab}
    \begin{tabular}{lcr}
        \hline
        Emission Line & Equivalent width \& Error & Width \\
        \hline
        Units & \AA & (kms$^{-1}$) \\
        \hline
        \ion{H}{i} 4340\AA\, (5-2) & -21.7$\pm$5.7 & 1746.6\\
        \ion{H}{i}4861\AA\, (4-2) & -49.3$\pm$2.6 & 2302.9\\
        \ion{H}{i}6562\AA\, (3-2) & -161.6$\pm$2.2 & 2747.5\\
        \ion{He}{i}5875\AA\, (3-2) & -31.3$\pm$4.2 & N/A\\
        \ion{He}{i}6678\AA\, (2-1) & -19.4$\pm$5.2 & N/A\\
        \ion{He}{i}7065\AA\, (1-0) & -16.2$\pm$4.9 & N/A\\
        \ion{He}{ii}4685.7\AA & -9.0$\pm$5.1 & N/A\\
        \hline
    \end{tabular}
\end{table}

\subsection{Light Curves}

\begin{figure}
	\includegraphics[width=\columnwidth]{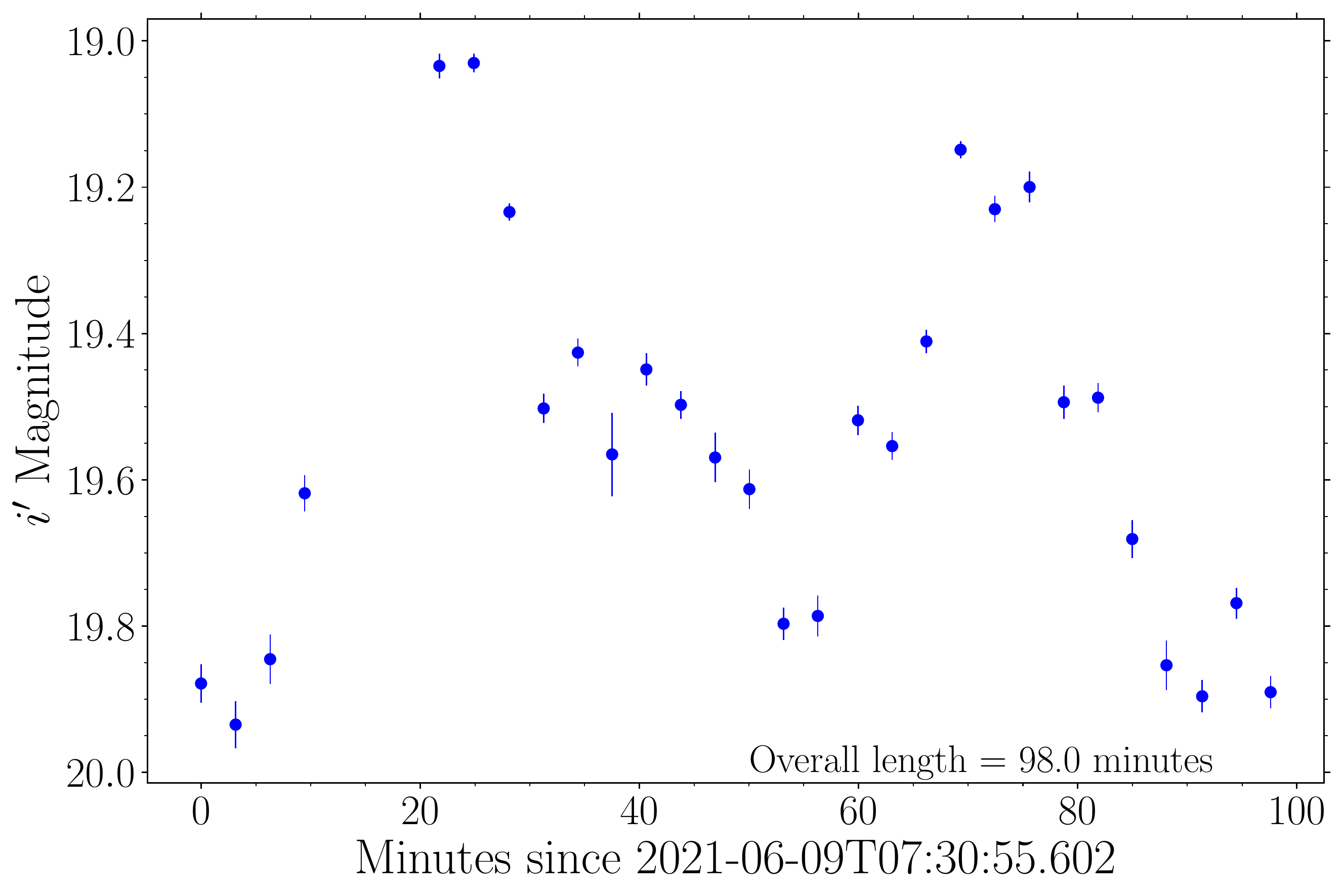}
    \caption{$i'$-Band light curve for Source 363, using SOAR's Goodman spectrograph in imaging mode. Fainter exposures were averaged with sequential epochs to recover signal. The peaks are separated by $\approx$51$\pm$2.2 minutes and the amplitudes are $\approx$0.9 mag.}
    \label{fig:curve_I}
\end{figure}

\begin{figure}
	\includegraphics[width=\columnwidth]{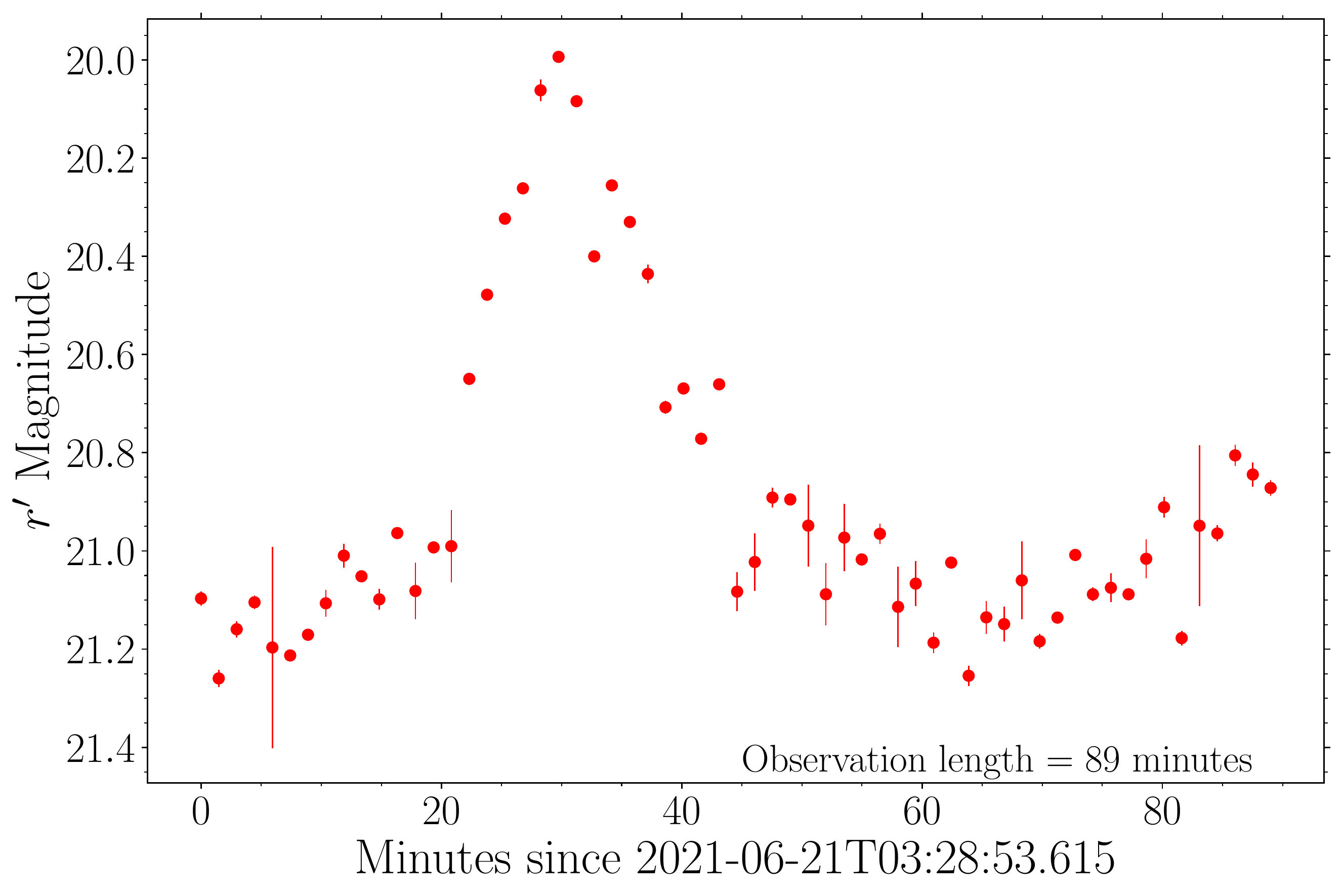}
    \caption{r'-band light curve, obtained with the PF-QHY instrument on the WHT. The peak shows an amplitude of $\approx$1.2 mag and resembles flare-like behaviour, with a comparatively fast rise followed by extended cooling time. If we assume this behaviour repeats based upon data from other filters, the timescale is of the order of approximately 66 minutes, from the start of the event to the next rise.}
    \label{fig:curve_r}
\end{figure}

\begin{figure}
	\includegraphics[width=\columnwidth]{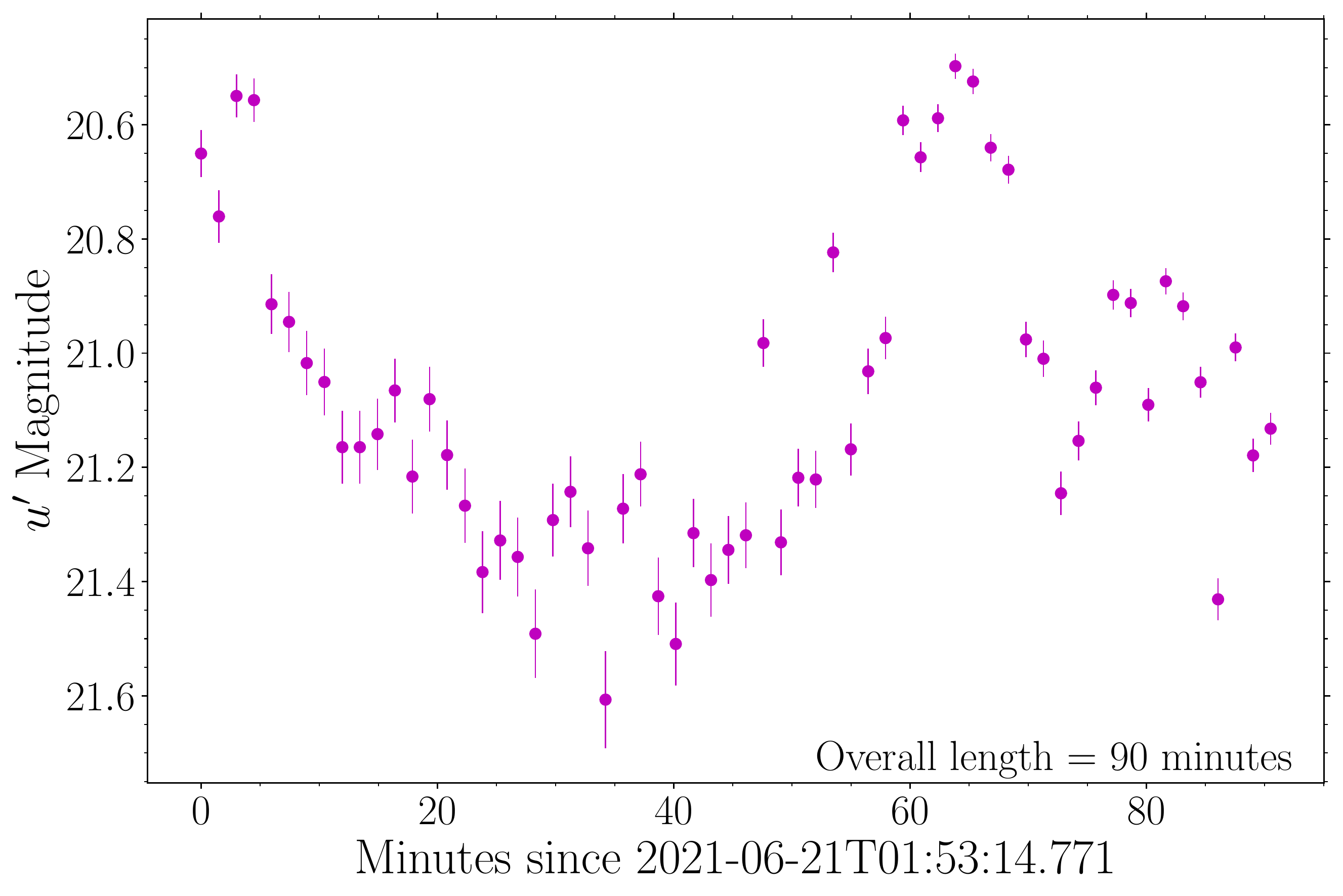}
    \caption{u'-band light curve, obtained with the PF-QHY instrument on the WHT. There is significant scatter in this curve, likely because of the u' bands increased sensitivity to changes in seeing. However the the same high amplitude, short timescale behaviour can be seen as in other filters. The two apparent peaks are separated by between 64 and 67 minutes, which is analogous to the timescales seen in the \textit{r'} curve.} 
    
    \label{fig:curve_u}
\end{figure}

The three light curves (Figures \ref{fig:curve_I}, \ref{fig:curve_r} and  \ref{fig:curve_u} for $u'$, $r'$ and $i'$ respectively) each display amplitudes between 0.9 and 1.2 mag, but each wavelength shows differing behaviour: The $u'$ curve shows symmetric behaviour with minimal time spent at either bright or faint states, whereas the $r'$ curve is at a faint level for approximately two thirds of the observation, with the rise taking $\approx 7~$min, followed by a 20 min decline to a state of quiescence, resembling behaviour reminiscent of stream-fed accretion on to the WD. This is similar to that seen in polar-type magnetic CVs (mCVs). The longer duration observations in $i'$ show two peaks, with a separation and shape consistent with the $u'$ data, including the scatter about the curves' minima. The peaks are separated by $51\pm2.2$~minutes and $65\pm2.1$~minutes for \textit{i'} and \textit{u'} respectively, with the \textit{r'} curve having a less clear cycle, although the curves' minima are separated by $\approx 64\pm2.1$~minutes. These estimates provide a rough baseline for the most visible timescales of variability.

The $u'$ and $i'$ light curves show small amplitude variation at low brightness states not seen at the maxima, although at this time it cannot be wholly rejected that these are indicators of short period variability (not unlikely given the magnetohydrodynamic processes present in the disc). We investigated this by fitting a third order spline to the phase-folded light curve (using \texttt{scipy}, \citealt{2020SciPy-NMeth}) and checking the residuals to this fit. This revealed no obvious periodicity. It would seem therefore that the variations during the system's minima are more likely the result of increased sensitivity to the atmospheric conditions during the observations. In addition a 19 hour LC (with approximately 1 hour on-source) in UVOT U-band was produced, consisting of 6 epochs of varying exposure time. Whilst not much can be gleaned from the shape (given the known rate of variability), the maximum amplitude was $\approx$0.8 mag, in keeping with data from other bands.  

Interpreting the nature of the observed variability is a challenge due to the limited time span of the observations, although a few causes can be inferred. A common cause of variability in CVs is flickering, a phenomenon that is thought to relate to modulation of the accretion rate and flares associated therein (see \citealt{2021Bruch} for a comprehensive review). A second driver of variability could be the spin of the WD primary star, seen via the accretion columns or polar hot-spots (depending on observed wavelength). 
A complication can arise from the orbital side-band: a side-band represents the reprocessing of the WD primary's emission by its disc, which causes an interaction between the spin period and the orbital period of the secondary, observed as a distinct frequency,
$\omega_{spin} \pm \omega_{orb}$, with a positive sign in most in cases, e.g. AO Psc \citep{2015AO_Psc} and a negative sign if the orbit is retrograde.
While there are currently not enough individual cycles to confirm any periods, we note that the lack of observed periodicities at short timescales ($\approx$20 min) in any of the curves is interesting. This could imply that the WD does not spin at the common period of approximately 10-15 min.

\subsection{X-ray Photometry}

The Swift-XRT observations detect a single source at RA = 295.7935$^{\circ}$, and Dec = 18.6475$^{\circ}$ with an astrometric uncertainty of 3.0\arcsec~ (see figure 7) given by the Swift-XRT data products generator’ software \citep{2014Evans_xrt}. These coordinates are offset from the optical/IR coordinates by 1.9\arcsec, which is consistent with zero offset within the uncertainties. We note that the rather scattered distribution of X-ray photons shown in Figure 7 is typical of Swift/XRT observations of a single faint source, see \url{https://www.swift.ac.uk/analysis/xrt/xrtcentroid.php} for an example of how real sources are distinguished from noise.

Given the lower resolution of Swift's instruments than the optical ones discussed thus far, the possibility of chance association should be considered. The measured X-ray flux is brighter than the ROSAT sensitivity limit (see below). Therefore we consider X-ray source density of 0.27 objects per square degree in the Galactic plane, as determined from the ROSAT catalogue \cite{ROSAT_Cat}) at Galactic coordinates $30^{\circ}<l<90^{\circ}$, $-3^{\circ}<b<3^{\circ}$. With this figure the probability of a chance alignment within 3.0\arcsec~ is $3.78 \times 10^{-6}$. Given that the fitted XRT spectrum is brightest at the lower energy regime, the low sensitivity of ROSAT to higher energies is not detrimental to this estimation. Further to this, whilst there are two optical/IR sources within the XRT astrometric uncertainty radius, the associated UVOT light curve displayed high amplitude variability consistent with that observed in Source 363 and the second star is not optically variable. Therefore it is clear that the XRT X-ray source is Source 363.

The average count rate over the length of observations was 0.01375 counts~$s^{-1}$; this was then converted into a flux via fitting an absorption adjusted power law to the XRT spectrum (see section \ref{swift_obs}), with a model temperature of 20~keV. The previous lack of detections implied that the overall flux would be low, but our measured fluxes (adjusted for absorption) of $1.3\substack{+0.4 \\ -0.3}\times10^{-12}\ \text{~erg~s}^{-1}\text{cm}^{-2}$ are above the lower limits for surveys such as ROSAT, making the non-detection interesting. Luminosities corresponding to these fluxes have a range that's higher than average for IPs. Depending on the distance used, luminosity ranges from $6.9\pm1.8\times10^{31}\ \text{~erg~s}^{-1}$ to $2.3\pm0.6\times10^{32}\ \text{~erg~s}^{-1}$, at the closest and furthest distance estimates respectively (as discussed in Section \ref{Source_info}). The associated X-ray light curve shows only an intermittent signal, implying that the source has a more extreme bright to faint state ratio than other similar stars. 

\begin{figure}
	\includegraphics[width=\columnwidth]{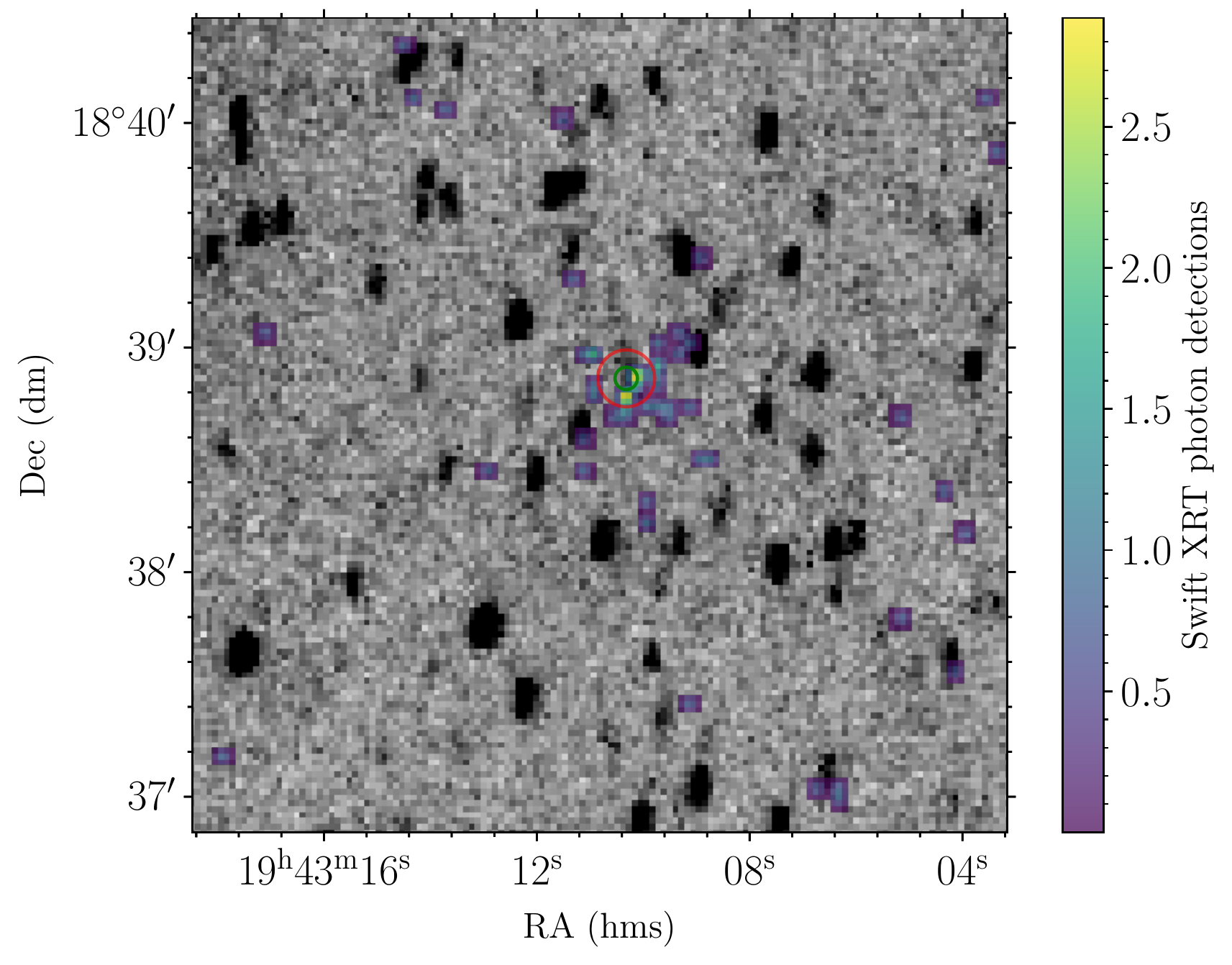}
    \caption{XRT photon-count mode image (blue/green/yellow) overlaid on a stacked UVOT-U band image (grayscale), Source 363 is located within the red (7.6\arcsec) and green (3.0\arcsec) annuli. These represent the source detection confidence radii (without and with UVOT astrometry respectively), as calculated by 'The Swift-XRT data products generator' \citep{2014Evans_xrt}. Our system is faint in the stacked UV image, but the XRT image provides the first indication that the star is an X-ray source. } 
    
    \label{fig:Swift_img}
\end{figure}

\section{Discussion}

\subsection{Comparison to sources in the literature} 

\begin{figure}
	\includegraphics[width=\columnwidth]{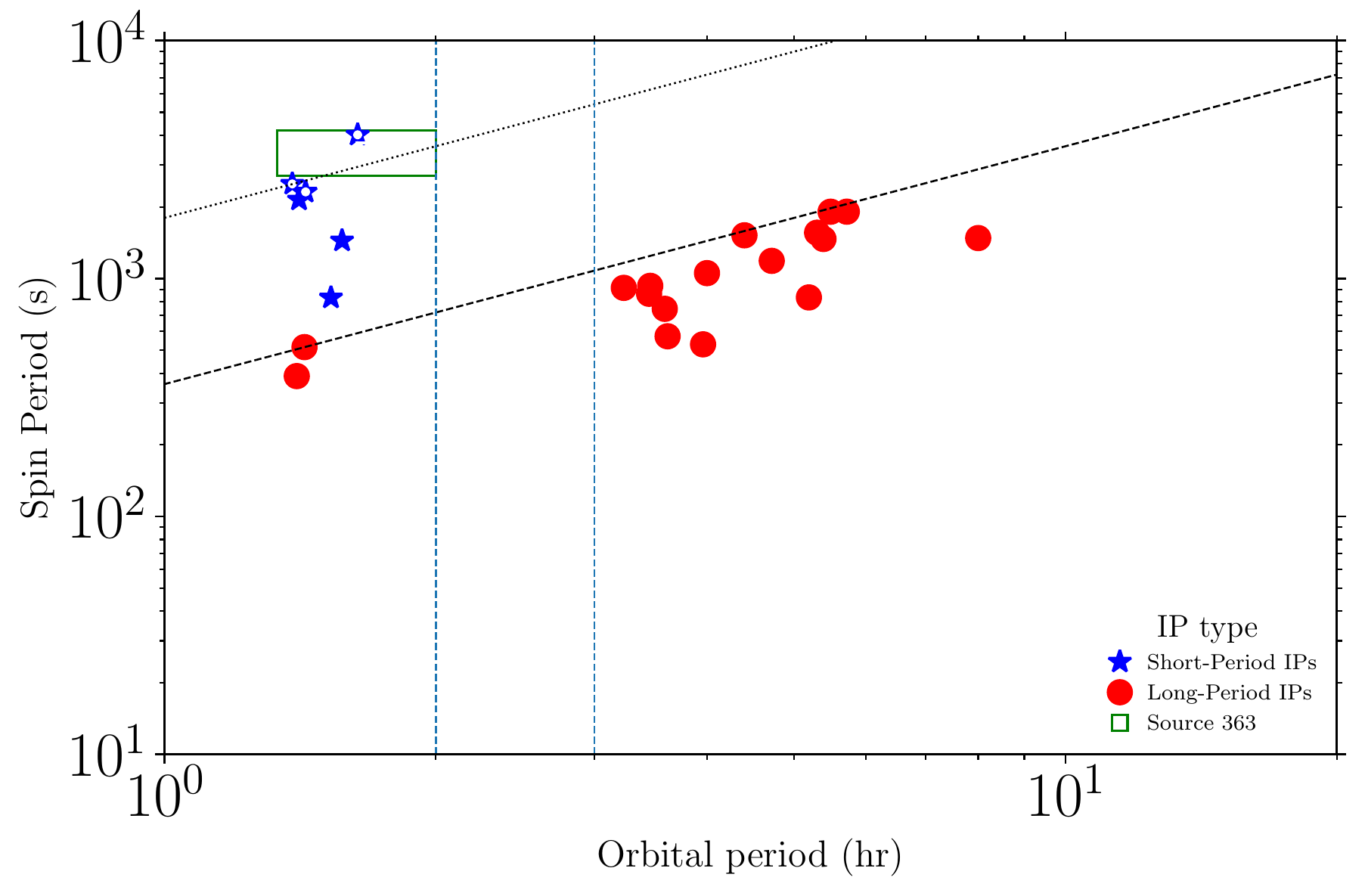}
    \caption{Orbital period versus spin period for a selection of known IPs (data from Mukai's 'The Intermediate Polars' - see footnote). The blue stars represent short period systems (those with white dots being mentioned in the text), red circles typical IPs. The green rectangle represents the likely range of orbital and spin periods of Source 363, these being constrained by the minimum orbital period of $\approx$80 mins \citep{Knigge2011} and the lower limit of the period-gap. The two dashed vertical lines at $P_{orb} = 2~hr~ \&~ 3~hr$ represent the 'period gap' where magnetic braking due to angular momentum transfer will cut-off mass-transfer between the WD and its companion. The diagonal dashed line is where $P_{orb} = 10 \times P_{spin}$, below which long-period IPs are found. The additional dotted line demarcates the region where the orbital period is twice the spin period, which is a common feature of short-period IPs.}
    \label{fig:mukai_repo}
\end{figure}

\begin{figure}
	\includegraphics[width=\columnwidth]{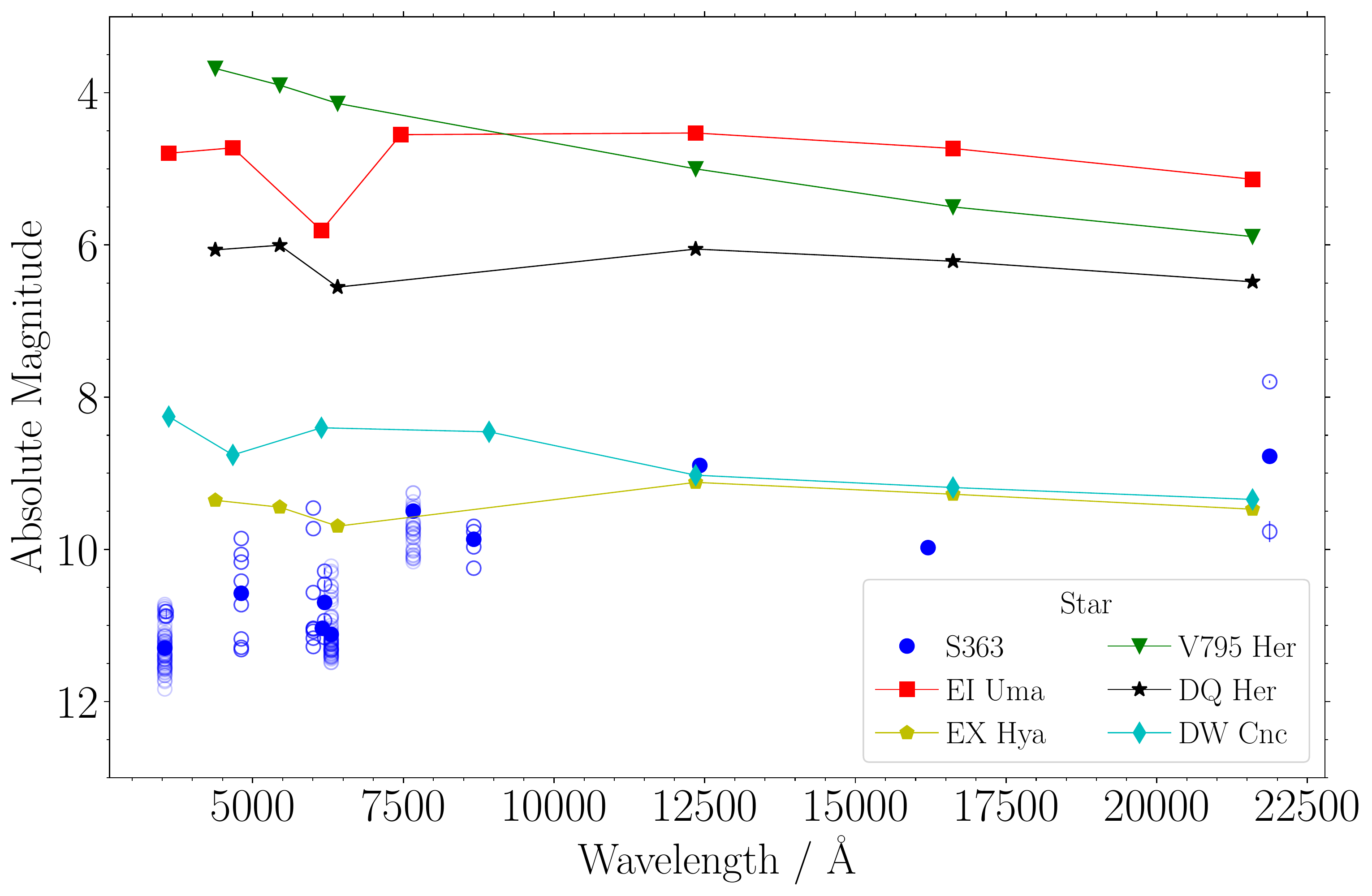}
    \caption{Optical and near infrared SEDs (AB system) for Source 363 (blue) and a selection of existing IPs from the literature. The solid blue points represent the median optical values from the new light curves in this work ($u'$, $r'$, $i'$), the 3 Panstarrs passbands and two IGAPS passbands with multiple epochs: ($g$,$r$,$z$) and ($u$, $r$), respectively. The single-epoch $J$ and $H$ UKIDSS data and the mean of the two UKIDSS $K$ values are also shown as solid blue points. Open circles are the individual measurements from which the medians were computed. (For readability the Panstarrs $r$, WHT $r'$ and IGAPS $r$ data in slightly different filters have been offset by 250~\AA but $u$ and $u'$ are almost identical so they are combined into one average). Note the high red optical and IR luminosity compared to shorter wavelengths. This is not present in previously classified IPs. The selected IPs are those with long spin periods and orbital periods ($P_{\mathrm{spin}}$~$\sim$~$0.3P_{\mathrm{orb}}$), with distances coming from \citep{Bailer_Jones2021}, using Gaia EDR3.}
    \label{fig:seds}
\end{figure}

Existing IPs with some degree of similarity to Source 363 have often been detected either through their X-ray emission or optical narrow-band flux. Thus we must compare the overall behaviour of Source 363 to these systems, in order to understand the observed behaviour. Constructing some approximate optical to near infrared spectral energy distributions (SEDs) (see Figure \ref{fig:seds}) we note that Source 363 has both the lowest overall luminosity at the wavelengths measured and an unexpectedly faint optical SED. This SED has been formed by averaging all available measurements at each wavelength (including each individual measurement from this work, as well as all available archive data), with the full range of values also marked on the plot. It should be noted that the \textit{J} (1.25~$\mu$m) and \textit{H} (1.65~$\mu$m) data are single measurements that may not be representative: the $J$ datum may be near a peak value, given the exceptionally blue $J-H$ colour (see $\S1$). Having noted this caveat, the SED appears to have a redder slope than other systems, possibly a byproduct of the disk being optically thicker than other examples. The two most similar objects (EX Hya and DW Cnc) are both short-period systems that lack discs, and exist in a separate portion of the IP spin-orbit period diagram, where $P_{orb} \approx 2 \times P_{spin}$ (see Figure \ref{fig:mukai_repo} \footnote{References for Figure \ref{fig:mukai_repo}: \cite{Hilton2009}, \cite{Rodriguez-Gil2004}, \cite{2013Andronov}, \cite{1998Buckley_V1025,1989ABuckley_TXCol}, \cite{1993Hellier_TVCol}, \cite{2002Kemp_HTCam}, \cite{2002Norton_V2306}, \cite{2003Staude_MUCam}, \cite{2003Woudt_YYSex,2012Woudt_CCScl},\cite{2005Kim_BGCmi}, \cite{2011Joshi_WXPyx}, \cite{2012Masetti_1RXS}, \cite{1991Mateo_YYDra}, \cite{1988Kal_Sem_AOPisc}, \cite{1996Burwitz_UUCam}, \cite{1997Hellier_PQGem,2006Evans_PQGem}, \cite{2002Norton_V2306}, \cite{2009Pretorius_IGRJ17165}, \cite{2008Sazonov_IGRJ08} - The URL of the Mukai's 'The Intermediate Polars' (where this data is collated) is as follows: '\url{https://asd.gsfc.nasa.gov/Koji.Mukai/iphome/iphome.html}'}), as opposed to the more common ratio $P_{orb} \approx 10 \times P_{spin}$.

We propose that our source might also be a part of this small population, given the lack of a detectable short spin period in any of our observations, and the orbital period lower limit given the low mass of the companion (see Section 4.2). Our $u'$ light curve is also similar in shape to DW Cnc \citep{Rodriguez-Gil2004}, although it is a factor 2 higher in amplitude, with both sources having periods of under an hour. The corresponding $r'$ light curve from the same night of observations shares little in common with any typical IP system in the literature.

A more comparable object is the short-period IP V598 Peg (SDSS~J233325.92+152222.1, \citealt{Southworth2007} \& \citealt{Szkody2005}), which has a spin period of 43 minutes, and a confirmed (via XMM-Newton direct observations by \citealt{Hilton2009}) orbital period of 86 minutes. The shape of the optical light curve when phase folded (Figure 4 in Hilton et al.) is almost identical to the r' band curve (see Figure\ref{fig:curve_r}).  In addition, the H$\alpha$ to H$\beta$ ratio of 1.87 is of a similar order to Source 363's.

To compare our spectrum to the literature, we find \citet{Oliveira2017} to be a good source of similar candidates due to their survey also using SOAR's Goodman spectrograph. Source 363's slightly blue continuum (by comparison to polar-like sources) and broad lines bear the closest resemblance to MLS2054-19, MLS0720+17, CSS1012-18 and SSS1359-39 (see Figure 1 in the aforementioned paper). Interestingly, three of these are classified by those authors as discless IPs (MLS0720+17 was suggested to be a polar). Some small differences to these sources are worth noting, in particular the relative strength of the H$\beta$ line compared to H$\alpha$, with all of these sources having the 4-2 transition as strong or stronger than the 3-2 line, the inverse of our finding.       


\subsection{Nature of the System?}

As it stands, we can say with some confidence that if the {\bf apparent} $\sim$45-70 minute variation timescales are related to the WD spin period (e.g. with the spread of values caused by stochastic flickering) then the system cannot be a polar CV. All hydrogen-rich stars will over-fill, rather than fill, their Roche lobes for such short periods (leading to a rapid cut off of the mass transfer via magnetic braking), and the optical spectrum of the source clearly shows strong hydrogen emission lines, indicating that accretion is taking place. The argument could be made that there are two accretion streams, from both poles of the WD being magnetized (synchronising the system) much like the system V2400 Oph \citep{Hellier02}, and thus the spin period would then be on the order of 90 minutes. However given that the spectrum suggests a disc, it is more likely the orbital period is longer than the spin period and just below `period gap', making the system some form of IP. 

With the IP explanation preferred based on the optical spectrum, it then becomes important to investigate the evolutionary state of the system, in order to explain the unusual behaviour (high amplitude, red SED and low overall luminosity). We can add constraints to the orbital period of the system by considering its near-infrared brightness and the amplitude of the associated variability; using the Bailer-Jones Gaia EDR3 distance of 900.7$\substack{+288.3 \\ -245.6}$~pc, we find Source 363 has $M_{K} \approx 6$, and in order to explain the observed variability the companion object must be be at least one magnitude fainter than that. This is because the variability is intrinsic to the WD, and thus the companion must not be providing the majority of the flux. These two factors combined lead us to place constraints on the classification of the companion, finding that it should be no more massive than an M3V star at 0.36~$\text{M}_\odot$ \citep{2013Mamajek}. This proposed limit on the companion mass would put the orbital period at roughly 2.5 hours or less, where a 2.5 hr period sits in the IP `period gap'. Equally using Figure 13 in \citet{Knigge2011} for our two \textit{Ks} observations leads to orbital period limits of between 2.1 and 1.4 hours. While this is a wide spread of values it does place the system below the period gap, in the region between that and the minimum IP orbital period of 80 minutes. 

Thus we can reasonably then assume the orbital period is less than the 2.1 hour lower limit \citep{1982rappa}, especially given the higher than average X-ray luminosity (stars within the gap have very low accretion rates, and thus low X-ray luminosity). Additionally a companion star of the described nature would have a short orbital period if in an IP-like system, lending further credence for Source 363 to exist below the period-gap. There is an argument to be made that the orbital period is in fact less than 90 minutes, as can be seen in Figure \ref{fig:mukai_repo}, because a number of short-period IPs (including the aforementioned V598 Peg) have settled into orbital periods at twice that of the white dwarf spin, but with the same magnetic moment as more normal IPs, as predicted in \citet{Webbink+Wick02}. With an orbital period in this range we would expect a typical IP system to have a spin period on the order of 5-9 minutes, which is not seen in our data, hence we posit that Source 363 may be a member of the short period sub-class mentioned above. This would roughly place the spin period of the white dwarf between 41 and 63 minutes, which mimics the timescales observed in our optical data.

Finally the observed strong IR excess (as described in Section 4.1) can then be assumed to be tracing the accretion flow, made more likely due to the lack of cyclotron emission features in the optical spectrum. Combining this with the short period, we can infer that our system has a reasonably high accretion rate for its short-period type. The method for this accretion is still uncertain, but is could be novel; Our light curves suggest direct accretion, whilst our SED and spectra suggest the presence of a disc, and thus disc-fed accretion. In order to make sense of this dichotomy, there is a possible explanation that Source 363 is currently undergoing an evolutionary transition between a typical (long-period) IP, and the short-period systems, such as has been described by \cite{Southworth2007} and \cite{2008Norton}.

There is currently an open question as to what the number density of short-period IPs should be \citep{2014Pret_Mukai}. The small binary separations in short period systems should easily allow the WD magnetic fields to connect with the convective donor star, and thus quickly synchronise the orbits, with system thus becoming a polar CV - provided $\mu_{WD} \lesssim 5 \times 10^{33}\ \mathrm{Gcm}^{3}$ \citep{NWS04}. This interpretation would render the subgroup short-lived, thus explaining their low observed numbers. On the other hand white dwarf magnetic fields can fall over time (Ohmic decay), and hence dramatically extend the time taken to become synchronous, in turn making such systems quite old, increasing the number density of this subgroup by extension. With a higher accretion rate than usual for short-period IPs, ram pressure would likely exceed magnetic pressure, which would then produce a disc, lending credence to this explanation for Source 363.

\section{Conclusions}

We report on the discovery of a new magnetic cataclysmic variable star system with an observed near-infrared excess, and have determined that our source is most likely a short-period IP system, with a higher than average accretion rate. This decision was based upon the following: 
\begin{itemize}
    \item Our optical spectrum shows emission of \ion{H}{i}, \ion{He}{i} and \ion{He}{ii}, all of which are indicators of magnetic CV systems. In addition the presence of disc-like features indicates that the magnetic field is not solely responsible for the accretion, making the case that the star is a polar unlikely. 
    
    \item The low optical luminosity of the system implies a low mass companion to the WD, which sets an upper limit on the orbital period of the binary. 

    \item Combining this with the lack of an obvious spin period near the expected range of 5 to 9 minutes, we can infer that the system does not fit with the standard 1 : 10 ratio between spin and orbital timescales. This deduction then places Source 363 within the loose group of short-period IPs, such as EX Hya. 
    
    \item The confirmed detection of X-ray emission from the system in targeted observations is interesting given that the star is absent from previous wide field X-ray surveys that cover the region (it does not feature in \citealt{Pretorius2013}). The non-detection by ROSAT would normally suggest that the overall X-ray luminosity is lower than typical IPs, a phenomenon that has been observed in other short-period systems. However, the measured X-ray luminosity of $1.31\pm0.34\times10^{32}\text{~erg~s}^{-1}$ (using the Gaia EDR3 Bailer-Jones estimated distance of 900~pc) is higher than that of most short period IPs and we would expect the source to have been detected by ROSAT, if the luminosity were constant. We can then infer that this system goes through bright phases, relative to an otherwise faint X-ray flux level.    
    
    \item Additionally, we make a case that given the short orbital period and low optical brightness, that the system has a lower than average accretion rate (for an IP in general) and a companion star of a very faint type. This provides contrast with other short-period systems, making Source 363 seem fairly unique. 
\end{itemize}

With such an unusual nature, implications from this work are twofold: 
\begin{itemize}
    \item It re-iterates that there is a population of under-luminous CVs  that continue be missed in all wide area X-ray surveys completed thus far. But this work indicates a chance that the X-ray luminosity of these sources may not be as low as current literature suggests. Previous short-period systems have $\text{L}_{X}\approx10^{31}\text{~erg~s}^{-1}$ whereas Source 363 ($L_X \approx 10^{32}$~erg~s$^{-1}$) shows that these systems can exist above this cut but still below the more usual value of $\approx10^{33}\text{~erg~s}^{-1}$ seen in typical IP systems. With the suggestions of a population of under-luminous IPs that may be responsible for some of the observed X-ray excess in the galactic bulge \citep{2016Hailey_GBXS}, Source 363 might be a useful laboratory to probe this idea further. The ongoing eROSITA \citep{eROSITA} survey should be deep enough to detect similar objects. 
    \item It shows the usefulness of using near infrared variability to locate this kind of object, as well as the benefits of multi-wavelength astronomy to classify variable stars that are unexpected or unusual upon initial discovery. 
\end{itemize}

The next steps in our investigation are to further constrain the system's parameters, specifically the WD and companion masses, as well as the spin and orbital periods. In addition, having these parameters will allow us to test our ideas for the system's unusual behaviour, as described in Section 4.2.
In addition we aim to find additional candidates within the Vista Variables in the Via Lactea survey (VVV, \citealt{Minniti2010}), via collation of short period variable stars \citep{virac} with close distances and observed H\textsubscript{$\alpha$} excess from VPHAS+ \citep{vphas}. This will allow us to build an idea of the scale of this population, and thus the expected X-ray flux that would be produced. In turn, this would provide a useful metric to determine whether they are the cause of the aforementioned flux excess. Further the results thereof would indicate whether we are looking at a new subgroup of IPs or a short-lived phase in their evolution.

\section*{Acknowledgements}
 

We acknowledge the use of public data from the Swift data archive. We also acknowledge the use of the WHT: The William Herschel Telescope and its service programme are operated on the island of La Palma by the Isaac Newton Group of Telescopes in the Spanish Observatorio del Roque de los Muchachos of the Instituto de Astrofísica de Canarias. We would also like to thank the operations teams at both facilities for the quick and efficient observations.

CM acknowledges support from the UK's Science and Technology Facilities Council (ST/S505419/1).
NM and WJC are funded by University of Hertfordshire studentships; furthermore CM, PWL, NM, WJC, ZG and JED recognise the computing infrastructure provided via STFC grant ST/R000905/1 at the University of Hertfordshire.

JS acknowledges support from the Packard Foundation and National Science Foundation grant AST-1714825. Portions of this work were performed while SJS held a NRC Research Associateship award at the Naval Research Laboratory. Work at the Naval Research Laboratory is supported by NASA DPR S-15633-Y.

ZG acknowledges the financial support from ANID (CONICYT) through the FONDECYT project No. 3220029. 

\section*{Data Availability}

All data is available from the first author upon request (and from PWL at p.w.lucas@herts.ac.uk), it will also be hosted at (star.herts.ac.uk/\textasciitilde cmorris/source\_363\_public/). Swift ToO data is publicly available at HEASARC.  



\bibliographystyle{mnras}
\bibliography{main_paper} 

\bsp	
\label{lastpage}
\end{document}